\documentclass[12pt]{article}
\usepackage{graphicx}
\usepackage{amssymb}
\usepackage{amsmath}
\newcommand{\MeV}{\mbox{~MeV}}

\newcommand{\ie}{{\it i.e.}}
\newcommand{\eg}{{\it e.g.}}
\newcommand{\eqn}[1]{&\hspace{-0.6em}#1\hspace{-0.6em}&}
\begin{document}
\baselineskip 0.6cm
%
\begin{titlepage}
\begin{center}

\begin{flushright}
\end{flushright}

\vskip 2cm

{\Large \bf 
Lepton Universality in the $\nu$MSM}

\vskip 1.2cm

{\large 
Takehiko Asaka$^1$,
Shintaro Eijima$^2$, and Kazuhiro Takeda$^3$
}

\vskip 0.4cm

$^1${\em
  Department of Physics, Niigata University, Niigata 950-2181, Japan
}

$^2${\em
  Institut de Th\'eorie des Ph\'enom\`enes Physiques, \'Ecole Polytechnique
F\'ed\'erale de Lausanne, CH-1015 Lausanne, Switzerland
}

$^3${\em
  Graduate School of Science and Technology, Niigata University, Niigata 950-2181, Japan
}

\vskip 0.2cm

(October 2, 2014)

\vskip 2cm

\vskip .5in
\begin{abstract}
  We consider the $\nu$MSM which is an extension of the Standard Model
  by three right-handed neutrinos with masses below the electroweak
  scale, in which the origins of neutrino masses, dark matter, and
  baryon asymmetry of the universe are simultaneously explained.
  Among three heavy neutral leptons, $N_2$ and $N_3$, which are
  responsible to the seesaw mechanism of active neutrino masses and
  the baryogenesis via flavor oscillation, can induce sizable
  contributions to various lepton universality in decays of charged
  mesons.  Then the possible deviations of the
  universality in the $\nu$MSM are investigated.  We find that the deviation in kaon
  decay can be as large as ${\cal O}(10^{-3})$, which will be probed in
  near future experiments.
\end{abstract}
\end{center}
\end{titlepage}
\renewcommand{\thefootnote}{\#\arabic{footnote}} 
\setcounter{footnote}{0}
%
\section{Introduction}
The $\nu$MSM (neutrino Minimal Standard
Model)~\cite{Asaka:2005an,Asaka:2005pn} is a simple extension of the
Standard Model (SM), explaining the origins of neutrino masses, dark
matter and baryon asymmetry of the universe at the same time.  Three
right-handed neutrinos are introduced with Majorana masses below the
electroweak scale ${\cal O}(100)$~GeV, which realize the seesaw
mechanism~\cite{Seesaw} for neutrino masses with very suppressed
Yukawa couplings.  The model predicts three heavy neutral leptons
$N_I$ ($I=1,2,3$) in addition to ordinary active neutrinos $\nu_i$
($i=1,2,3$).

The lightest heavy neutral lepton $N_1$ with ${\cal O}(10)$~keV mass
is a candidate for dark matter (see, for example, a
review~\cite{Boyarsky:2009ix}).  The others $N_2$ and $N_3$ with
quasi-degenerate masses can generate baryon asymmetry of the universe
through the mechanism given in~\cite{Akhmedov:1998qx,Asaka:2005pn}.
Enough baryon asymmetry can be generated even if the degenerate mass
$M_N$ of $N_2$ and $N_3$ is as small as ${\cal
  O}(1)$~MeV~\cite{Canetti:2010aw,Asaka:2013jfa}.  However, the lower
bound on masses is further restricted to avoid constraints from direct
searches and cosmology~\cite{Gorbunov:2007ak}.  The recent
analysis~\cite{Asaka:2013jfa} shows that $M_N > 163$~MeV for the
normal hierarchy (NH), while $M_N = 188-269$~MeV and $M_N>285$~MeV for
the inverted hierarchy (IH) of active neutrino masses.  It is
remarkable that, thanks to the smallness of masses, the heavy neutral
leptons in the $\nu$MSM, especially $N_2$ and $N_3$, can be directly
tested by a variety of experiments and/or
observations~\cite{Kusenko:2004qc,Gorbunov:2007ak,Atre:2009rg,Asaka:2012hc,Asaka:2012bb}.

These heavy neutral leptons mix with flavor neutrinos and their mixing
elements are given by the ratios between Dirac and Majorana masses.
It is then possible to produce $N_I$ by decays of various mesons
through the mixing as the production of ordinary active neutrinos.  As
an example, when they are sufficiently lighter than charged
kaon, the decays $K^+ \to e^+ N_I$ and $K^+ \to \mu^+ N_I$ are
possible.  In fact, these channels are good targets for direct search
of heavy neutral leptons by using the technique of the so-called peak
search experiment~\cite{Shrock:1980vy}.

Furthermore, such decays may spoil lepton universality of charged
meson decay~\cite{Shrock:1980ct,Shrock:1981wq}. For instance, it is
possible that the ratio of decay rates ($M = \pi, K, \cdots$)
\begin{eqnarray}
  \label{eq:LU}
  R_M 
  = \frac{\Gamma (M^+ \to e^+ \nu)}{\Gamma (M^+ \to \mu^+ \nu)} \,,
\end{eqnarray}
is significantly different from the SM prediction.  Although each
partial decay width receives considerable hadronic uncertainties, the
theoretical prediction can be very precise by taking the ratio, and
thus $R_M$ offers a promising test for physics beyond the SM. The
general expression for the contribution to $R_M$ from heavy neutral
leptons had already been presented in Ref.~\cite{Shrock:1980ct}.

Recently, Refs.~\cite{Abada:2012mc,Abada:2013aba} had revisited the
importance of this issue and violations of various universality
including $R_M$ had been extensively studied.  Especially, the
numerical estimation of $R_M$ in the inverse seesaw model had been
performed.  In addition, they had also pointed out that $R_M$ can be applied
 in the $\nu$MSM.

In this letter, following these developments, we estimate the
possible deviation of $R_M$ induced by heavy neutral leptons in the
$\nu$MSM.  The deviation strongly depends on masses and mixing
elements of $N_I$.  The mixing elements $\Theta_{\alpha 1}$ of $N_1$
must be very suppressed in order to avoid various constraints of dark
matter.  We then find that the contribution of $N_1$ to $R_M$ is
negligible.  Thus, $N_2$ and $N_3$ for the seesaw mechanism and
baryogenesis give the dominant contributions to lepton universality.

The main purpose of this letter is to identify the possible deviations
of lepton universality in the $\nu$MSM.  Hereafter, we first summarize
the constraints on heavy neutral leptons $N_2$ and $N_3$ and present
the allowed region of their mixing elements in Sec.~2.  
We then consider in Sec.~3 lepton universality in decays of light
mesons, $R_K$ and $R_\pi$, in the $\nu$MSM and estimate the
deviations from the SM.  Current status and future
perspective of experiments of lepton universality are also discussed.
Finally, Sec.~4 is devoted to conclusion.

\section{Heavy Neutral Leptons in the $\nu$MSM}
First of all, we explain briefly the $\nu$MSM.  Three right-handed
neutrinos are introduced with Lagrangian
\begin{eqnarray}
  {\cal L}
  = 
  i \overline{\nu_{R I}} \gamma^\mu \partial_\mu \nu_{R I}
  - F_{\alpha I} \overline {L_\alpha} \Phi \nu_{R I}
  - \frac{M_I}{2} \overline{\nu_{R I}^c} \nu_{R I} + h.c. \
\end{eqnarray}
Here and hereafter, we follow the notation presented in
Ref.~\cite{Asaka:2011pb}.  The seesaw mechanism works when Dirac
masses $F_{\alpha I} \langle \Phi \rangle$ are much smaller than
Majorana masses $M_I$.  In this case mass eigenstates of neutrinos are
three active neutrinos $\nu_i$ with masses $m_i$ and three heavy
neutral leptons $N_I$ with masses $M_I$.  Then, the neutrino mixing is
given by
\begin{eqnarray}
  \nu_{L \alpha}
  = U_{\alpha i} \, \nu_i + \Theta_{\alpha I} \, N_I^c \,,
\end{eqnarray}
where $U_{\alpha i}$ are elements of the PMNS
matrix~\cite{Pontecorvo:1958,Maki:1962mu}, and $\Theta_{\alpha I} =
F_{\alpha I} \langle \Phi \rangle/M_I$ are mixing elements of heavy
neutral leptons.

Heavy neutral lepton $N_1$ with $M_1 = {\cal O}(10)$~keV plays a role
of dark matter.  The mixing elements of $N_1$ must be suppressed
enough since too large $|\Theta_{\alpha 1}|$ would lead to the
overclosure of the universe due to too much present abundance and also
would provide too much X-rays from its radiative decay $N_1 \to \nu
\gamma$%
\footnote{Recently, the unidentified line spectrum is observed
  \cite{Bulbul:2014sua,Boyarsky:2014jta,Boyarsky:2014ska}, which can
  be interpreted by X-ray lines emitted by sterile neutrino dark
  matter (\ie, $N_1$ in the considering model).  }
(see Ref.~\cite{Boyarsky:2009ix}).  It is then found that $N_1$ can
only give negligible contribution to the seesaw mass matrix of active
neutrinos and can essentially play no role in baryogenesis to avoid
these difficulties.  In addition, as will be discussed later, $N_1$
contribution to the ratio $R_M$ in Eq.~(\ref{eq:LU}) can be neglected
compared with those from $N_2$ and $N_3$.  Therefore, we shall take
$|\Theta_{\alpha 1}| = 0$ in this analysis for simplicity.

Heavy neutral leptons $N_2$ and $N_3$ are then responsible to the mass
matrix for active neutrinos via the seesaw mechanism and also the
baryogenesis via flavor oscillation.  In this case, to realize the
seesaw mechanism Yukawa coupling constants $F_{\alpha I}$ of $N_2$ and
$N_3$ can be expressed as follows~\cite{Casas:2001sr}:
\begin{eqnarray}
  \label{eq:FP}
  F_{\alpha I} = \frac{i}{\langle \Phi \rangle} \, 
  \left[ U \, D_{\nu}^{\frac{1}{2}} \,
  \Omega \, D_N^{\frac{1}{2}} \right]_{\alpha I}\,.
\end{eqnarray}
Here and hereafter we shall follow the notation in
Ref.~\cite{Asaka:2011pb}: $U$ represents the PMNS matrix, 
\begin{eqnarray}
 U=\left( 
\begin{array}{ccc}
c_{12}c_{13} & s_{12}c_{13} & s_{13}e^{-i\delta} \\
-c_{23}s_{12}-s_{23}c_{12}s_{13}e^{i\delta}& c_{23}c_{12}-s_{23}s_{12}s_{13}e^{i\delta} & s_{23}c_{13} \\
 s_{23}s_{12}-c_{23}c_{12}s_{13}e^{i\delta}& -s_{23}c_{12}-c_{23}s_{12}s_{13}e^{i\delta}  & c_{23}c_{13} \\
\end{array} 
\right) \times {\rm diag}(1, \ e^{i\eta}, \ 1 ),
\end{eqnarray}
with $s_{ij}= \sin \theta _{ij}$, $c_{ij}= \cos \theta _{ij}$, 
$D_{\nu} ={\rm diag}(m_1,m_2,m_3)$ and $D_N ={\rm diag}(M_2, M_3)$. The matrix
$\Omega$ is given by
\begin{eqnarray}
  \Omega = \left(
    \begin{array}{ccc}
      0 & 0 \\
      \cos \omega & -\sin \omega \\
      \xi \sin \omega&\xi \cos \omega \\
    \end{array} 
  \right) \mbox{for the NH case}\,,~~  
  \Omega = \left(
    \begin{array}{ccc}
      \cos \omega & -\sin \omega \\
      \xi \sin \omega& \xi \cos \omega \\
      0 & 0 \\
    \end{array} 
  \right)
  \mbox{\rm for the IH case}\,.
\end{eqnarray}
The couplings are written in terms of parameters of active neutrinos
and heavy neutral leptons.  The former ones consist of masses $m_i$ as
well as mixing angles $\theta_{ij}$, Dirac phase $\delta$ and Majorana
phase $\eta$ in the PMNS matrix.%
\footnote{ Since $N_1$ essentially decouples from the seesaw
  mechanism, the lightest active neutrino obtains a mass smaller than
  ${\cal O}(10^{-5})$ eV~\cite{Asaka:2005an}.  The number of Majorana
  phases in the PMNS matrix is effectively reduced to be one (rather
  than two in the usual case with three massive active neutrinos).  }
The latter ones are a complex parameter $\omega$, masses $M_{2,3}$ and
the sign parameter $\xi$.  As for the masses, the successful
baryogenesis requires that $N_2$ and $N_3$ are quasi-degenerate in
mass, and we write them in the form $M_3 = M_N + \Delta M/2$ and $M_2
= M_N - \Delta M/2$ with $\Delta M \ll M_N$.
The imaginary part of $\omega$ is important to determine
the typical size of the mixing elements since
$|\Theta_{\alpha I}| \propto X_\omega \equiv \exp (\mbox{Im}\omega)$.
In fact, as shown in Ref.~\cite{Asaka:2011pb},
$|\Theta_{\alpha I}|$ can be large 
by taking $X_\omega \gg 1$ without changing masses of active neutrinos.

The mixing elements $\Theta_{\alpha I}$ characterize the strength of
interactions for heavy neutral leptons, and then receive constraints
from direct searches and cosmology.  Interestingly, as pointed out in
Ref.~\cite{Gorbunov:2007ak}, the former ones place the upper bounds on
$|\Theta_{\alpha I}|$ while the latter one gives the upper bound on
lifetimes $\tau_{N_2}$ and $\tau_{N_3}$ leading to the lower bounds on
$|\Theta_{\alpha I}|$.  Consequently, we may obtain the allowed range
of the mixing elements.  Such regions have already been evaluated in
Refs.~\cite{Gorbunov:2007ak,Asaka:2013jfa}.  Here we reconsider this
issue, especially taking into account for the first time 
the preliminary result from the BNL-E949
experiment~\cite{E949:2014}.  Notice that we shall restrict ourselves
for the case when $M_N < m_K - m_e$ because such heavy neutral
leptons, as we will show later, induce a significant deviation of the
lepton universality in kaon decay.

In deriving the allowed region, we construct Yukawa couplings of $N_2$
and $N_3$ by using the central values of $\theta_{ij}$ and $\Delta
m_{ij}^2$ from the global analysis of neutrino oscillations in
Ref.~\cite{Forero:2014bxa} and by varying all the possible ranges for
other free parameters.  We then show the allowed range for the
combination of $\Theta_{\alpha I}$
\begin{eqnarray}
  |\Theta|^2 \equiv  \sum_{I=2,3} \sum_{\alpha = e, \mu, \tau} 
  |\Theta_{\alpha I}|^2 \,,
\end{eqnarray}
for a given $M_N$.%
\footnote{ The mass difference $\Delta M$ gives negligible corrections
  to all the results in the present analysis, and hence we take
  $\Delta M = 0$ for simplicity.  } In our parameterization of Yukawa
couplings, it is written as
\begin{eqnarray}
  | \Theta |^2  = \frac{ \sum_{i=1,2,3}m_i}{2 M_N} 
  (X_\omega^2 + X_\omega^{-2}) \,.
\end{eqnarray}

As for the bounds from direct search experiments, we first consider
the case when $M_N < 450$~MeV and use the results from the peak search
experiments~\cite{E104,Britton:1992xv,PIENU:2011aa,E949:2014} as well
as the beam-dump
experiments~\cite{Bernardi:1985ny,Bernardi:1987ek,Levy:1986w}.  (See
the discussion later for the case in which $m_K - m_e > M_N >
450$~MeV.)  Following Refs.~\cite{Kusenko:2004qc,Ruchayskiy:2011aa}
we have taken into account the corrections applying the bounds from
PS191 experiment~\cite{Bernardi:1985ny,Bernardi:1987ek,Levy:1986w} to
the $\nu$MSM, \ie, the targets are two heavy neutral leptons $N_2$ and
$N_3$ which are Majorana particles (the target is one Dirac particle in the
original analysis), and the neutral current contributions for decays
of heavy neutral leptons are added (such a contribution is neglected in the
original analysis).  

Moreover, the successful baryogenesis also gives the upper bounds on
the mixing elements in order to avoid the strong washout of the
produced asymmetry~\cite{Canetti:2010aw}.   However, as shown in Ref.~\cite{Canetti:2010aw}, such bounds are much weaker than those from PS191 experiment in the considering mass range.

In this analysis, we also consider the recent
bound from BNL-E949 experiment~\cite{E949:2014}.  It is the peak
search experiment in $K^+ \to \mu^+ \nu$ decay giving the upper bound
on $|\Theta_{\mu I}|^2$.  Finally, to avoid the cosmological
difficulty we impose the lifetime bound $\tau_{N_{2,3}} < 0.1$
s~\cite{Dolgov:2000pj,Dolgov:2000jw}.  Unfortunately, the analysis
in Refs.~\cite{Dolgov:2000pj,Dolgov:2000jw} has been done in the
different situation from the $\nu$MSM.  We then also discuss the case
when the lifetime bound is relaxed as $\tau_{N_{2,3}} < 1$ s to make
the most conservative analysis. To evaluate $\tau _{N_{2,3}}$, 
we use the formulae of the partial decay widths of heavy neutral leptons
given in Ref.~\cite{Gorbunov:2007ak}.

\begin{figure}[t]
  \centerline{
  \includegraphics[width=8cm]{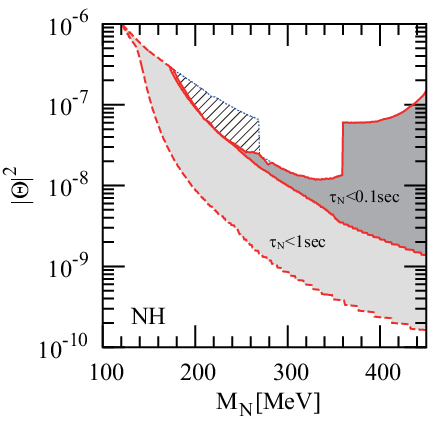}%
  \includegraphics[width=8cm]{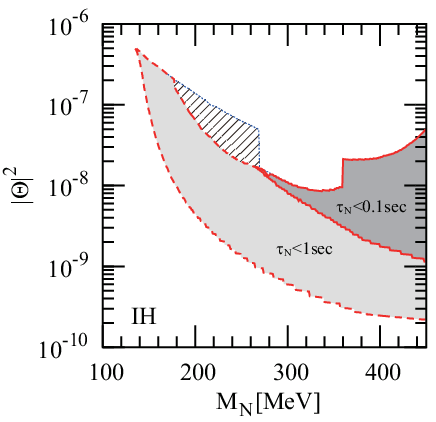}%
  }%
  \caption{\it
    Allowed region in the $M_N$-$|\Theta|^2$ plane for the NH
    case (left panel) and IH case (right panel).
    Allowed regions are shown by the shaded regions with 
    red-solid line or red-dashed line for the case with
    the cosmological lifetime bound $\tau_{N_{2,3}}< 0.1$ s
    or $\tau_{N_{2,3}}< 1$ s, respectively. The hatched regions are excluded by the 
    bounds from BNL-E949 experiment~\cite{E949:2014}.  
  }
  \label{fig:AR_MN_THsq_1sec}
\end{figure}
The results are summarized in Fig.~\ref{fig:AR_MN_THsq_1sec}.  We find
that BNL-E949 experiment gives the more stringent bound for 
$M_N \simeq 180-260$~MeV compared with the bounds from PS191
experiment, which is seen by the hatched regions in Fig. 1.
  (See also the result in Ref.~\cite{Asaka:2013jfa} for
comparison.%
\footnote{
  Ref.~\cite{Asaka:2013jfa}
  had used the data of global analysis of the neutrino oscillations in Ref.~\cite{Fogli:2012ua} 
  rather than Ref.~\cite{Forero:2014bxa} used in this analysis. 
}) Especially, in the IH case, the lower bound on $M_N$
changes a lot by the inclusion of such a bound.  We then find that the
allowed mass region when $\tau_{N_{2,3}} < 0.1$ s is%
\footnote{ For the NH case the small mass region $M_N = 208$--211~MeV
  is excluded.  }
\begin{eqnarray}
  M_N > 
  \left\{
    \begin{array}{l l}
      173 \MeV & ~~\mbox{for the NH case}
      \\
      264\MeV & ~~\mbox{for the IH case}
    \end{array}
  \right. \,.
\end{eqnarray}
It should be noted that, if the cosmological upper bound of the
lifetime is relaxed as $\tau_{N_{2,3}} < 1$ s, the lower bound on
$M_N$ becomes smaller as $M_N > 122$~MeV and 136~MeV for the NH and IH
cases, respectively.  See also Fig.~\ref{fig:AR_MN_THsq_1sec}.
Therefore, the cosmological bound on the lifetime is crucial 
for determining the lower bound of the masses of $N_2$ and $N_3$.%
\footnote{
  The lifetime bound for the case when $M_N < m_\pi$ had also been
  discussed in Ref.~\cite{Ruchayskiy:2012si} and had shown that the
  mass region $M_N < m_\pi$ is excluded.  To make a very conservative
  analysis, however, we also consider the case where the lifetime of
  $N_{2,3}$ is longer than the limit in \cite{Ruchayskiy:2012si}.
} 

It is seen that the allowed range in Fig.~\ref{fig:AR_MN_THsq_1sec}
is very limited for both NH and IH cases.  
In practice, all such regions can be verified if the sensitivity
of $|\Theta|^2$ by future experiments will be improved by 
a factor of ${\cal O}(10^2)$ or ${\cal O}(10^3)$ 
when applying the lifetime bound $\tau_{N_{2,3}}< $ 0.1
or 1~s, respectively.  Such experiments will be
not only the peak search and beam-dump experiments,
but also the precision measurements of lepton universality
of light meson decays as shown below.

\section{Lepton Universality in the $\nu$MSM}
Let us discuss lepton universality of charged meson decays 
shown in Eq.~(\ref{eq:LU}) in the context of the $\nu$MSM.
We first consider the universality in 
charged kaon decay $R_K$.
The SM prediction of $R_K$ is 
\begin{eqnarray}
  R_K^{\rm SM} = 
  \left( \frac{m_e}{m_\mu} \right)^2 
  \left( \frac{ m_K^2 - m_e^2} {m_K^2 - m_\mu^2} \right)^2 
  \left( 1 + \delta R_K \right) 
  \,,
\end{eqnarray}
where $\delta R_K$ denotes the radiative correction.
Notice that $K^+ \to e^+ \nu_e$ and $K^+ \to \mu^+ \nu_\mu$ occur through
charged current interaction and their rates are
helicity-suppressed.  It should be mentioned that
both decay rates receive the hadronic uncertainties,
\eg, through the decay constant of parent meson, such uncertainties
cancel to a large extent by taking the ratio.
The theoretical prediction of the SM is thus very 
precise as~\cite{Finkemeier:1995gi,Cirigliano:2007xi}
\begin{eqnarray}
  R_K^{\rm SM } = (2.477 \pm 0.001) \times 10^{-5} \,.
\end{eqnarray}
In addition, the measurements at high precision have been done~\cite{Ambrosino:2009aa,Goudzovski:2010uk,NA62:2011aa,Lazzeroni:2012cx}.
The recent NA62 experiment provides~\cite{Lazzeroni:2012cx}
\begin{eqnarray}
  R_K^{\rm exp} = (2.488 \pm 0.010) \times 10^{-5} \,.
\end{eqnarray}
It is seen that the observational data agrees with
the SM value at the 1$\sigma$ level.
Consequently, the deviation
\begin{eqnarray}
  \Delta r_K = \frac{R_K}{R_K^{\rm SM}} - 1 \,,
\end{eqnarray}
is as small as
\begin{eqnarray}
  \Delta r_K = (4 \pm 4) \times 10^{-3} \,,
\end{eqnarray}
and thus it provides a powerful probe for physics beyond the SM.

In the $\nu$MSM, $K^+$ is possible to decay into not only active
neutrinos $\nu_i$ but also heavy neutral leptons $N_I$
depending on $M_I$.  Then, the ratio $R_K$ is given by
\begin{eqnarray}
  R_K \eqn{=} 
  \frac{
    \sum_{i=1,2,3} \Gamma (K^+ \to e^+ \nu_i)
    + 
    \sum_{I=1,2,3} \Gamma (K^+ \to e^+ N_I)
  }{
    \sum_{i=1,2,3} \Gamma (K^+ \to \mu^+ \nu_i)
    + 
    \sum_{I=1,2,3} \Gamma (K^+ \to \mu^+ N_I)
  }  \,.
\end{eqnarray}
The general expression of $R_M$ in the presence of heavy neutral
leptons has been given by Ref.~\cite{Shrock:1980ct}. (See Eq.~(3.2) in
Ref.~\cite{Shrock:1980ct}.)  By neglecting the masses of active
neutrinos and the experimental energy thresholds of charged leptons in
kaon decays, the deviation is~\cite{Shrock:1980ct}
\begin{eqnarray}
  \Delta r_K
  \eqn{=}
  \frac{
    \sum_{i=1,2,3} |U_{e i}|^2 +
    \sum_{I=1,2,3} |\Theta_{e I}|^2 G_{e I}
  }{
    \sum_{i=1,2,3} |U_{\mu i}|^2 +
    \sum_{I=1,2,3} |\Theta_{\mu I}|^2 G_{\mu I}
  }
  - 1 \,,
\label{eq:DelrK0}
\end{eqnarray}
where $G_{\alpha I} = 0$ if $M_I > m_K - m_{\ell_\alpha}$; otherwise
\begin{eqnarray}
  G_{\alpha I} =
  \frac{ r_\alpha + r_I - (r_\alpha - r_I)^2 }
  { r_\alpha ( 1 - r_\alpha)^2}
  \sqrt{ 1 - 2 (r_\alpha + r_I) +(r_\alpha - r_I)^2 } \,,
\end{eqnarray}
with $r_\alpha = m_{\ell_\alpha}^2/m_K^2$ and $r_I = M_I^2 /m_K^2$. 
(See Ref.~\cite{Gorbunov:2007ak} for the expressions of $\Gamma(K^+\rightarrow 
l_{\alpha}^+N_I)$.)

The physical importance of $\Delta r_K$ (and also $\Delta r_\pi$ in
the later discussion) had been readdressed in
Refs.~\cite{Abada:2012mc,Abada:2013aba}.  The main origins of such
deviations are (i) the additional contributions to the kaon decay from
heavy neutral leptons and (ii) the deviation from the unitarity of the
PMNS mixing matrix of active
neutrinos~\cite{Shrock:1980ct,Shrock:1981wq,Abada:2012mc,Abada:2013aba}.
Refs.~\cite{Abada:2012mc,Abada:2013aba} had presented the possible
range of $\Delta r_{K, \pi}$ in the inverse seesaw model and also had
pointed out that $\Delta r_{K, \pi}$ in Eq.~(\ref{eq:DelrK0}) can be
applied to the $\nu$MSM.

Based on these analyses, we would like to derive the predicted range
of $\Delta r_K$ in the $\nu$MSM.
First of all, it should be noted that 
the mixing elements of active neutrinos and heavy neutral leptons
satisfy the unitarity condition
\begin{eqnarray}
  \label{eq:Unitarity}
  \sum_{i=1,2,3} |U_{\alpha i}|^2
  + \sum_{I=1,2,3} |\Theta_{\alpha I}|^2 = 1 \,.
\end{eqnarray}
It is seen that the violation of the unitarity in the PMNS matrix $U$ 
is very suppressed at ${\cal O}(|\Theta_{\alpha I}|^2)$
in this framework (see Fig.~\ref{fig:AR_MN_THsq_1sec}).
From the above condition
$\Delta r_K$ in Eq.~(\ref{eq:DelrK0}) can be written as
\begin{eqnarray}
  \Delta r_K
  \eqn{=}
  \frac{
    1 + 
    \sum_{I=1,2,3} |\Theta_{e I}|^2 
    \left[ G_{e I} - 1 \right]
  }{
    1 +
    \sum_{I=1,2,3} |\Theta_{\mu I}|^2 
    \left[ G_{\mu I} - 1 \right]
  }
  - 1
  \,.
  \label{eq:delRk}
\end{eqnarray}
Therefore, we find that the deviation $\Delta r_K$ in the $\nu$MSM is
determined by the masses $M_I$ and mixing elements $\Theta_{\alpha I}$
of heavy neutral leptons.  
Note that $\Delta r_K$ does not depend
explicitly on the PMNS matrix elements, but it
depends on them implicitly through $\Theta_{\alpha I}$.
(See the parametrization of Yukawa couplings of $N_I$ in
Eq.~(\ref{eq:FP}).)  Since the mixing elements of dark matter
$N_1$ must be very small, we can safely neglect its contribution to
$\Delta r_K$.

First, we consider the case when $M_N < m_K - m_\mu$, \ie, both $K^+
\to \mu^+ N_I$ and $K^+ \to e^+ N_I$ are kinematically allowed.  In
this case one might expect that the deviation $\Delta r_K$ is very
suppressed as ${\cal O}(10^{-9})$--${\cal O}(10^{-7})$ since
$|\Theta|^2$ should be in such a range as shown in
Fig.~\ref{fig:AR_MN_THsq_1sec}.  Decay rate of $K^+ \to \ell_\alpha^+
N_I$ (and then $G_{\alpha I}$) is, however, enhanced by
$(M_I/m_{\ell_\alpha})^2$ compared with $K^+ \to \ell_\alpha^+
\nu_\alpha$ due to the helicity suppression~\cite{Shrock:1980ct}.  Interestingly, since
this enhancement factor is much larger for the decay into $e^+$ than
that into $\mu^+$, the $\nu$MSM predicts a positive $\Delta r_K$ in
this mass region as
\begin{eqnarray}
  \label{eq:DELRK_ap}
  \Delta r_K \simeq \sum_{I=2,3} | \Theta_{eI} |^2 \,
   \frac{M_N^2}{m_e^2}
  \left(
    1- \frac{M_N^2}{m_K^2} \right)^2 \,.
\end{eqnarray}
 Moreover, the upper limit of $\Delta r_K$ is then
derived from the upper bounds on the mixing elements $|\Theta_{e I}|$.
Such elements are severely restricted by PS191 experiment looking for
the production and decay modes $K^+ \to e^+ N_I$ and $N_I \to e^+
\pi^-, e^- \pi^+$, \eg, $|\Theta_{e I}|^2 < {\cal O}(10^{-9})$--${\cal
  O}(10^{-8})$ for $M_N \simeq 200$--400~MeV.  Therefore, we expect
$\Delta r_K <{\cal O}(10^{-4})$--${\cal O}(10^{-3})$ by taking into
account the enhancement factor of $(M_N/m_e)^2 \sim10^{5}$.

When $m_K - m_e >
M_N > m_K - m_\mu$, $K^+ \to \mu^+ N_I$ is forbidden, but the behavior
of the correction $\Delta r_K$ is very similar to the above case.
On the other hand, when $M_N > m_K - m_e$, the situation is changed.
We should note that, even if $K^+ \to \mu^+ N_I$
and $K^+ \to e^+ N_I$ are kinematically forbidden,
the correction of $\Delta r_K$ is induced due to the non-unitarity 
of the PMNS matrix (see Eq.~(\ref{eq:Unitarity})) as
\begin{eqnarray}
  \label{eq:DEL_rK_h}
  \Delta r_K \simeq \sum_{I = 2,3} 
  \bigl( |\Theta_{\mu I}|^2 - |\Theta_{e I}|^2 \bigr) \,.
\end{eqnarray}
In this case the sign of $\Delta r_K$ is determined according to the
relative sizes of $|\Theta_{\mu I}|^2$ and $|\Theta_{e I}|^2$ and the
magnitude is $|\Delta r_K |\lesssim |\Theta|^2 = {\cal O}(10^{-9})$--${\cal
  O}(10^{-7})$.

\begin{figure}[t]
  \centerline{
  \includegraphics[width=8cm]{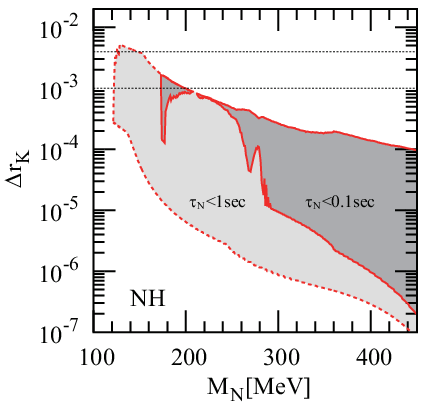}%
  \includegraphics[width=8cm]{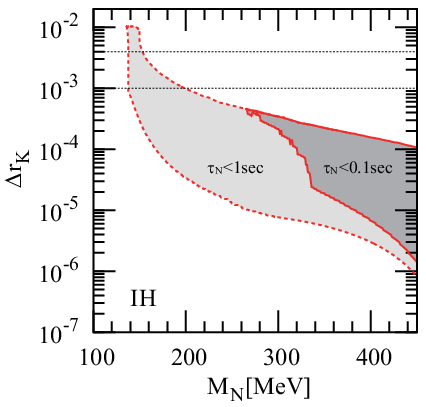}%
  }%
  \caption{\it
    $\Delta r_K$ in the $\nu$MSM for the NH case (left panel) 
    and IH case (right panel).
    Possible regions are shown by the shaded regions with 
    red-solid line or red-dashed line for the case with
    the cosmological lifetime bound $\tau_{N_{2,3}}< 0.1$ s
    or $\tau_{N_{2,3}}< 1$ s.  
    The horizontal (black dotted) lines are 
    $\Delta r_K = 4 \times 10^{-3}$ (current central value~\cite{Lazzeroni:2012cx})and 
    $\Delta r_K = 10^{-3}$ (which will be reached by
    the near future experiments).
  }
  \label{fig:DEL_RK_1sec}
\end{figure}

Now, we are at the point to present the numerical prediction of
$\Delta r_K$ in the $\nu$MSM.  As explained in Sec.~2, we impose the
constraints from direct search experiments and cosmological lifetime bound.
The possible range of $\Delta r_K$ by varying all the free parameters
is shown in Fig.~\ref{fig:DEL_RK_1sec}.  It is found that $\Delta r_K
={\cal O}(10^{-7})$--${\cal O}(10^{-3})$ for the NH case, and $\Delta
r_K = {\cal O}(10^{-6})$--${\cal O}(10^{-3})$ for the IH case, 
where we have considered $M_N < 450$ MeV and
$\tau_{N_{2,3}}<0.1$ s.  
The predicted region becomes wider if the
lifetime bound is relaxed as $\tau_{N_{2,3}}<1$ s.
The search bounds place the upper limit while the lifetime bound places
the lower limit of $\Delta r_K$, and hence the
$\nu$MSM predicts $\Delta r_K$ in certain range.  We find that the
predicted range is indeed consistent with the current upper bound at 3
$\sigma$ level, $\Delta r_K < 1.2 \times 10^{-2}$.

We have considered the mass range $M_N < 450$ MeV so far.  When $M_N >
450$ MeV, there is no stringent constraint on the mixing elements from
PS191 experiment.  So, we expect a large $\Delta r_K$ for $450$ MeV $<
M_N < m_K - m_e$.  In such a case, the upper bounds on
$|\Theta_{\alpha I}|$ are placed from CHARM and CHARM
II~\cite{Bergsma:1983rt, Bergsma:1985is, Vilain:1994vg},
IHEP-JINR~\cite{Baranov:1992vq} and NuTeV\cite{Vaitaitis:1999wq}
experiments.  When $M_N$ is just above 450~MeV, the most stringent
bound on $|\Theta_{eI}|^2$ is obtained from IHEP-JINR and the bound in
Fig.~4 of Ref.~\cite{Baranov:1992vq} is weaker than that of
PS191~\cite{Levy:1986w} by a factor of $\sim 40$.  This means that
$\Delta r_K$ can be $\sim 4 \times 10^{-3}$ in such a value of $M_N$.

In addition, the successful scenario of baryogenesis also puts the
important bound of the mixing elements for such mass regions.  We,
however, find that such a bound on $|\Theta|^2$ in
Ref.~\cite{Canetti:2012kh} is slightly weaker than the above IHEP-JINR
bound on $|\Theta_{eI}|^2$.  This shows that
search for heavy neutral leptons with $M_N$ just above 450~MeV
is very interesting since the present bounds from
beam-dump experiments, baryogenesis and also lepton universality
are very competitive and may be possible to be cross-checked
in various ways.
More precise estimation of these bounds as well as
$\Delta r_K$ in this case will be done elsewhere~\cite{AET}.

Near future experiments (such as NA62 at
CERN~\cite{Goudzovski:2012gh}, ORKA at
FNAL~\cite{E.T.WorcesterfortheORKA:2013cya} and TREK/E36 at
J-PARC~\cite{Kohl:2013rma}) will achieve the sensitivity $\Delta r_K =
10^{-3}$.  Therefore, it is very interesting that these experiments
will start to probe the predicted region in the $\nu$MSM.
In particular, large $\Delta r_K$ are obtained when $M_N \sim 180$ MeV
and just above 450 MeV.  Such mass regions will also be tested by
experiments using different search techniques, like
the peak search and/or beam-dump experiments,
in decays of kaon and charmed-mesons, respectively.

Next, we turn to consider lepton universality in pion decay.
The theoretical prediction of the SM is~\cite{Cirigliano:2007xi}
\begin{eqnarray}
  R_\pi^{\rm SM} = (1.2352 \pm 0.0001) \times 10^{-4} \,,
\end{eqnarray}
while the experimental value is~\cite{Beringer:1900zz}%
\footnote{ Here we have cited the averaged value of Particle Data
  Group~\cite{Beringer:1900zz}.  The recent measurements at TRIUMF and
  PSI give $R_\pi = (1.2265 \pm 0.0034 \pm 0.0044) \times
  10^{-4}$~\cite{Britton:1992pg} and $R_\pi= (1.2346 \pm 0.0035 \pm
  0.0036) \times 10^{-4}$~\cite{Czapek:1993kc}, respectively.  }
\begin{eqnarray}
  R_\pi^{\rm exp} = ( 1.230  \pm  0.004 ) \times 10^{-4} \,.
\end{eqnarray}
The deviation is then given as
\begin{eqnarray}
  \label{eq:DEL_rpi}
  \Delta r_\pi = ( -4 \pm 3 ) \times 10^{-3} \,.
\end{eqnarray}

\begin{figure}[t]
  \centerline{
  \includegraphics[width=8cm]{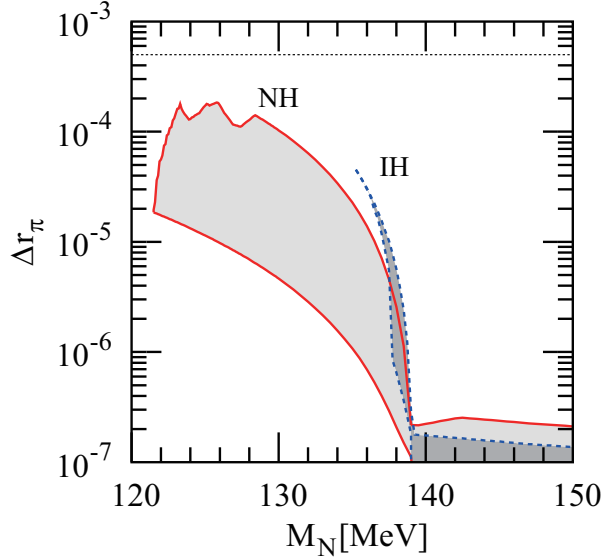}%
  }%
  \caption{\it $\Delta r_\pi$ in the $\nu$MSM.  Possible region is
    shown by the shaded region with red-solid line or blue-dashed
    line for the NH or IH case, respectively.  Here we impose the
    cosmological lifetime bound $\tau_{N_{2,3}}< 1$ s.  The
    horizontal (black dotted) line is $\Delta r_\pi = 5 \times
    10^{-4}$ (which will be reached by the near future experiments).
  }
  \label{fig:DEL_RP_1sec}
\end{figure}
In the considering model, $\pi^+ \to \mu^+ N_{2,3}$ are impossible even
if the lifetime bound is relaxed as $\tau_{N_{2,3}}< 1$ s and then
we restrict ourselves to the case with $M_N > m_\pi - m_\mu$.  When
$\pi^+ \to e^+ N_{2,3}$ are available, the sizable correction to
$R_\pi$ is expected due to the enhancement factor of $(M_N/m_e)^2$ and
its maximal value is determined by the upper bounds of
$|\Theta_{eI}|^2$.  Then, the approximate form of $\Delta r_\pi$
is given by  
\begin{eqnarray}
  \Delta r_\pi \simeq \sum_{I=2,3} | \Theta_{eI} |^2 \,
   \frac{M_N^2}{m_e^2}
  \left(
    1- \frac{M_N^2}{m_\pi^2} \right)^2 \,,
\end{eqnarray}
similar to Eq.~(\ref{eq:DELRK_ap}).
Moreover, the sign of $\Delta r_\pi$
is positive as in the kaon decay, and thus, even if $N_{2}$
and $N_3$ were allowed to be lighter than pion (to be precise $m_\pi -
m_e$), they would be conflict with $R_\pi^{\rm exp}$ at 1 $\sigma$
level (see Eq.~(\ref{eq:DEL_rpi})).  As shown in
Fig.~\ref{fig:DEL_RP_1sec}, we find numerically that the predicted
range is $\Delta r_\pi < {\cal O}(10^{-4})$, and then it is consistent
with $R_\pi^{\rm exp}$ at 2 $\sigma$ level.

Notice that the experiments like PIENU at
TRIUMF\cite{Malbrunot:2011zz} and PEN at PSI~\cite{Pocanic:2009zz}
will improve the sensitivity at the level $\Delta r_\pi \simeq 0.05
-0.06$ \% (see also Ref.~\cite{Bryman:2011zz}), which is slightly
above the predicted range.  Thus, the further improvement may be
required to probe $\Delta r_\pi$ in the $\nu$MSM.

When $M_N$ becomes larger than $m_\pi - m_e$, the non-unitarity of the
PMNS mixing matrix for active neutrinos induces the correction
\begin{eqnarray}
  \Delta r_\pi \simeq \sum_{I= 2,3}
  \bigl( |\Theta_{\mu I}|^2 - |\Theta_{e I}|^2 \bigr) \,,
\end{eqnarray}
as in the kaon decay (see Eq. (\ref{eq:DEL_rK_h})). In this case the magnitude
of $\Delta r_\pi$ is too small to be probed in near future
experiments.  It is, however, interesting to notice that $\Delta
r_\pi$ and $\Delta r_K$ become the same when $M_N > m_K - m_e$.

We have so far discussed the corrections to lepton universality in
kaon and pion decays.  It should be noted that heavy neutral leptons
$N_2$ and $N_3$ may lead to violations of lepton universality in
decays of charmed mesons, beauty mesons and tauon. See the recent 
analysis in Ref.~\cite{Abada:2013aba}. The comprehensive
study for the test of the $\nu$MSM by lepton universality will be
discussed elsewhere~\cite{AET}.

\section{Conclusions}
We have discussed lepton universality of charged meson decays in
the $\nu$MSM.  Among three heavy neutral leptons, $N_2$ and $N_3$,
which explain the seesaw mechanism for active neutrino masses and the
baryogenesis via their flavor oscillation, may induce the  violations
of such universality
due to the non-unitarity of the mixing matrix of active neutrinos and
the additional contributions to meson decays.

The deviation of lepton universality in kaon decay $R_K$ has been
found to be as large as $\Delta r_K ={\cal O}(10^{-3})$ when applying the
cosmological bound on lifetime as $\tau_{N_{2,3}}< 0.1$ s.  Such a
large $\Delta r_K$ is possible when $M_N \sim 180$ MeV and just above
$450$ MeV.  Further, if
the cosmological bound on the lifetime is weak as $\tau_{N_{2,3}}
\lesssim 1$ s, $\Delta r_K$ can be larger as ${\cal O}(10^{-2})$.
Notice that the sign of $\Delta r_K$ is always positive in the case
when $K^+ \to e^+ N_{2,3}$ are open.  Furthermore, we have also
discussed lepton universality in pion decay.  When $\pi^+ \to e^+
N_{2,3}$ are allowed by relaxing the lifetime bound, the deviation can
be as large as $\Delta r_\pi ={\cal O}(10^{-4})$.

Such regions of the model will begin to be explored by near future
experiments; the experiments of lepton universality in kaon decay as
NA62, ORKA and TREK/E36 experiments and those in pion decay as PIENU
and PEN experiments.  It should be noted that
such regions are also good targets of direct search experiments
using the different methods (the peak search experiments,
the beam-dump experiments, and so on).
These facilities might reveal physics of $N_2$
and $N_3$, namely the origins of neutrino masses and baryon asymmetry
of the universe.

\section*{Acknowledgments}
T.A. was supported by JSPS KAKENHI Grant Numbers 25400249 and 26105508, 
and S.E. was supported by Sinergia grant of the Swiss National Science Foundation CRSII2 141939.



\end{document}